\documentclass[amssymb,aps,prb,twocolumn,showpacs,superscriptaddress]{revtex4}

\usepackage{graphicx}
\usepackage{dcolumn}
\usepackage{bm}

\begin{document}
\title{Spin-polarized tunneling as a probe of (Ga,Mn)As electronic properties}
\author{M. Elsen}
\email{marc.elsen@paris7.jussieu.fr} \affiliation{Unit\'e Mixte de
Physique CNRS/Thales, Route Départementale 128, 91767 Palaiseau
Cedex, France and Universit\'e Paris-Sud 11, 91405 Orsay, France}
\author{H. Jaffr\`es}
\affiliation{Unit\'e Mixte de Physique CNRS/Thales, Route
Départementale 128, 91767 Palaiseau Cedex, France and Universit\'e
Paris-Sud 11, 91405 Orsay, France}
\author{R. Mattana}
\affiliation{Unit\'e Mixte de Physique CNRS/Thales, Route
Départementale 128, 91767 Palaiseau Cedex, France and Universit\'e
Paris-Sud 11, 91405 Orsay, France}
\author{L. Thevenard}
\affiliation{Laboratoire de Photonique et de Nanostructures,Route
de Nozay, 91460 Marcoussis, France}
\author{A. Lemaitre}
\affiliation{Laboratoire de Photonique et de Nanostructures,Route
de Nozay, 91460 Marcoussis, France}
\author{J.-M. George}
\affiliation{Unit\'e Mixte de Physique CNRS/Thales, Route
Départementale 128, 91767 Palaiseau Cedex, France and Universit\'e
Paris-Sud 11, 91405 Orsay, France}

\begin{abstract}
We present magnetic and tunnel transport properties of
(Ga,Mn)As/(In,Ga)As/(Ga,Mn)As structure before and after adequate
annealing procedure. The conjugate increase of magnetization and
tunnel magnetoresistance obtained after annealing is shown to be
associated to the increase of both exchange energy $\Delta$$_{exch}$
and hole concentration by reduction of the Mn interstitial atom in
the top magnetic electrode. Through a 6x6 band k.p model, we
established general phase diagrams of tunneling magnetoresistance
(TMR) and tunneling anisotropic magnetoresistance (TAMR) \textit{vs.}
(Ga,Mn)As Fermi energy (E$_F$) and spin-splitting parameter (B$_G$).
This allows to give a rough estimation of the exchange energy
$\Delta$$_{exch}$=6B$_G$$\simeq$120 meV and hole concentration
p$\simeq$1.10$^{20}$cm$^{-3}$ of (Ga,Mn)As and beyond gives the
general trend of TMR and TAMR \textit{vs.} the selected hole band
involved in the tunneling transport.
\end{abstract}

\date{\today}
\maketitle

\section{Introduction}
In the field of spintronics, the \textit{p}-type ferromagnetic
semiconductor (Ga,Mn)As offers many advantages to study tunnel
magnetotransport properties when used as an electrode. The complexity
of the transport mechanisms associated with spin-orbit coupled states
make this material a powerful means for finding novel effects and
provides new challenges for theoretical understandings. This includes
tunnel magnetoresistance (TMR) across single and double barriers,
\cite{TanakaPRL, MattanaPRL} tunnel anisotropic magnetoresistance
(TAMR), \cite{RusterPRL, SaitoPRL} Coulomb blockade anisotropic
magnetoresistance \cite{WunderlichPRL} and current induced
magnetization switching.\cite{ChibaPRL, ElsenPRB} However one of the
main limitation of this \textit{p} type material for spintronic
integration is the relatively low Curie temperature. Through low
temperature annealing treatement after growth, Curie temperatures of
173 K can be obtained. \cite{WangProc} Elimination of interstitial
manganese atoms which are double donors and which couple
antiferromagnetically with the manganese atom in substitutionnal
position, is mainly invoked. These atoms diffuse towards the surface
to form either a MnO\cite{EdmondsPRL, YuPRB}or a MnN\cite{KirbyPRB}
layer, depending on annealing conditions.\newline

In this paper we describe the effect of annealing on the magnetic and
electric properties of a (Ga,Mn)As/(In,Ga)As/(Ga,Mn)As tunnel
junction. We have focused our report on this single structure even
though other junctions with different Mn concentrations and
ferromagnetic layer thicknesses were studied, leading to the same
general conclusion. \cite{Elsenthesis} In the first part we detail
the effect of annealing on magnetization measurements and confirm
observations made on a (Ga,Mn)As trilayer structure with a GaAs
barrier.\cite{ChibaAPL} The second part presents the results obtained
on junctions fabricated by optical lithography and describes the
behaviour of Resistance Area (R.A) product, Tunnel Magnetoresistance
(TMR) and Tunnel Anisotropic Magnetoresistance (TAMR) through
annealing. In the last part, a general interpretation of the data
behaviour from both magnetic and electric measurements, is given
through a 6x6 band \textit{k.p} model of the tunneling transport. Two
important parameters are identified, the Fermi energy and the spin
splitting parameter B$_G$ introduced in the framework of the Zener
model in the mean-field approach.\cite{DietlPRB}

\section{Experimental results}

Ga$_{0.926}$Mn$_{0.074}$As (80nm)/ In$_{0.25}$Ga$_{0.75}$As (6nm)/
Ga$_{0.926}$Mn$_{0.074}$As (15nm) structure is grown by molecular
beam epitaxy at 250 $^o$C on a \textit{p}-doped GaAs buffer layer
(p$\cong$2$\cdot$10$^{19}$cm$^{-3}$). Annealing treatment has been
realized at 250 $^o$C in a nitrogen atmosphere during 1 hour. The
annealing was performed on a whole piece of 5x5 mm$^{2}$ for magnetic
measurements whereas realised on patterned junctions for
electrical experiments. \\

\begin{figure}
\includegraphics[width=8.5cm]{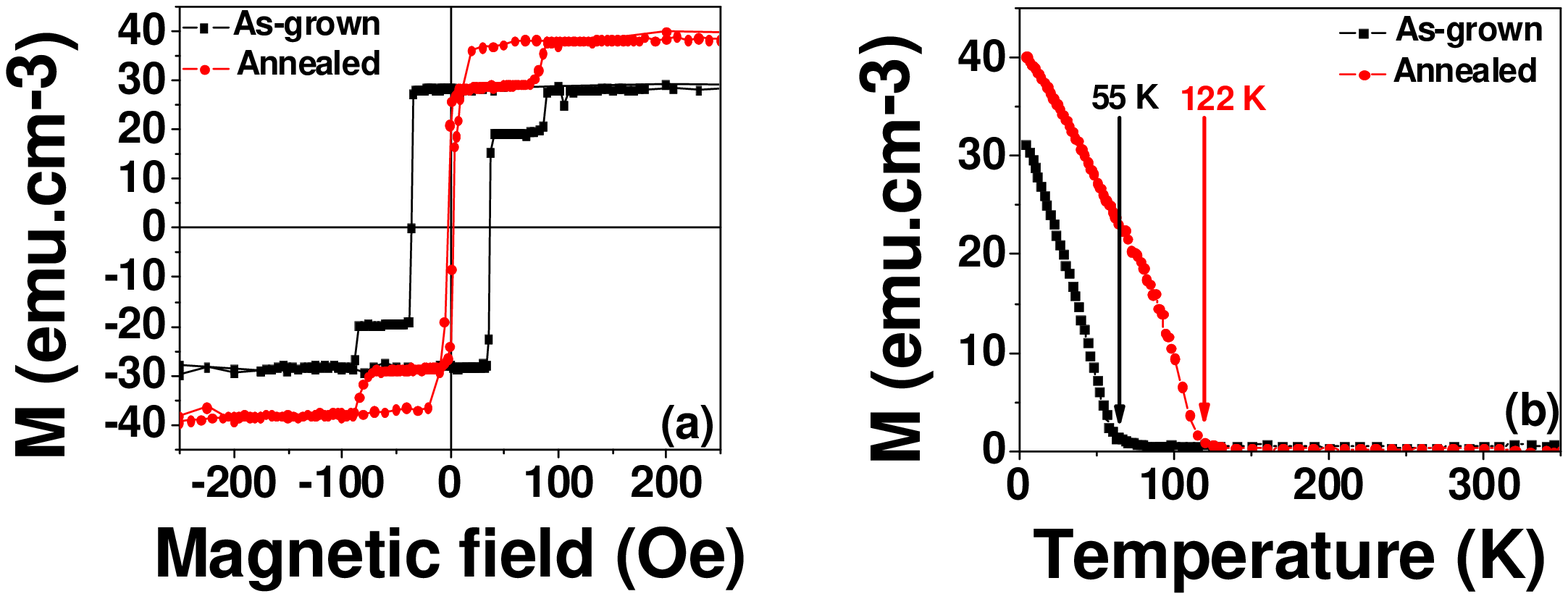}
\caption{ (Color online) (a) Magnetization measurements \textit{vs.}
magnetic field at 10 K before and after annealing along [100]
direction; (b) Magnetization measurements as a function of the
temperature before and after annealing in a field of 500 Oe.}
\label{Magnetization}
\end{figure}

Figure 1 presents magnetization behaviour before and after annealing
by SQUID (Superconducting Quantum Interference Device) measurements.
The two step magnetization reversal along [100] axis at 10 K is due
to the consecutive reversal of the two magnetic layers [Fig.
\ref{Magnetization}a]. As a function of annealing, three important
characteristics, consistent with the reduction of Mn interstitials in
the top magnetic layer, can be extracted from those measurements : a
large decrease of the coercivity H$_C$ as well as an increase of the
magnetic moment M$_S$ and of the Curie temperature T$_C$. Concerning
the variation of the H$_C$, magnetization study pointed out that this
decrease may be due to an elimination of interstitial manganese
(double donors) pinning center.\cite{PotashnikJAP} The resulting
increase of the carrier concentration may also contribute to the
decrease of the anisotropy field. \cite{DietlPRB} Considering that
only the top layer is affected by annealing, through its linear
dependence on the magnetization saturation value, the spin splitting
parameter increases from 17 meV (before annealing) to 24 meV (after
annealing). The values of the spin splitting were estimated through
the relationship $B_{G}$=$\frac{A_F \beta M_S}{6 g \mu_{B}}$ derived
from the mean field theory, where A$_F$ is the Fermi Liquid parameter
and $\beta$ the \textit{p-d} exchange integral.\cite{DietlPRB}
\newline

The observed Curie temperature are in good agreement with those found
on thicker magnetic layers confirming that layer width larger than 50
nm should still have a high concentration of manganese interstitials.
\cite{YuPRB,YuPRBbis} In the present case, the Curie temperature goes
from 55 K to 122 K [Fig. \ref{Magnetization}(b)]. Due to the higher
magnetic moment of the top layer, the behaviour of the thin bottom
layer is hidden, supporting the results that only the top layer
properties change. A further confirmation that annealing does not act
on the bottom layer comes from Auger measurements, not presented
here. A strong manganese accumulation at the top of the surface is
measured, whereas no obvious change in the bottom layer is observed,
already put forward by Chiba \textit{et al.} \cite{ChibaAPL} Capping
(Ga,Mn)As layer by a simple GaAs layer which width exceeds 5 nm, does
not improve the Curie temperature of the simple magnetic layer
\cite{StoneAPL} and does though support our results. The formation of
a \textit{p-n} junction avoiding the migration of interstitial
\textit{n}-type manganese has been suggested.\cite{KirbyAPL}
\newline

\begin{figure}
\includegraphics[width=8.5cm]{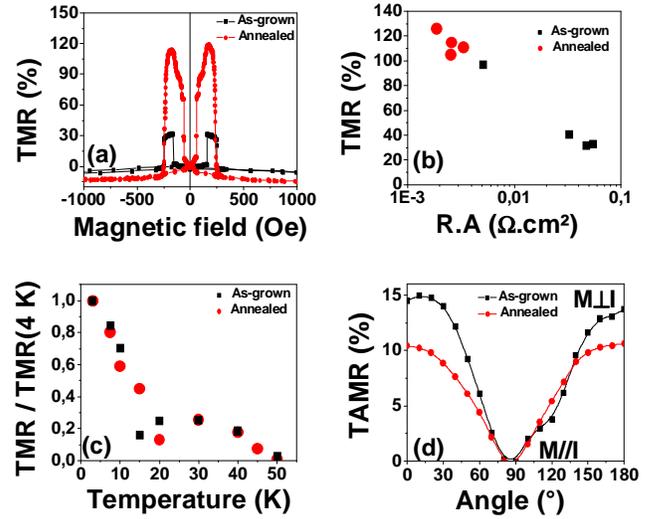}
\caption{ (Color online) (a) Tunnel magnetoresistance measurements as
a function of the magnetic field at 1 mV and 3K for a 128 $\mu$m$^2$
junction. (b) Tunnel magnetoresistance measurements as a function of
Resistance.Area product at 3 K for 4 (un)annealed junctions. (c)
Tunnel magnetoresistance at 1 mV as a function of the temperature
before and after annealing. (d) Tunnel anisotropic magnetoresistance
measurements as a function of the magnetic field at 1 mV and 3K.}
\label{Measurements}
\end{figure}

Magnetic tunnel junctions have been patterned by optical lithography
(size of the junctions were between 8 and 128 $\mu$m$^2$). With
standard dc technique the resistance of the junctions is measured at
3K and at low bias (1 mV) in the CPP (Current Perpendicular to Plane)
regime. Non-linear I(V) curve indicates that a 6 nm (In,Ga)As layer
still acts as a barrier.\cite{ElsenPRB} The reason is on the one hand
that the Mn acceptor level in GaAs leads to a positive band offset in
(Ga,Mn)As compared to GaAs. On the other hand the well-known As
antisites incorporated during the low growth process should probably
govern the pinning of the Fermi level and a higher barrier than the
simple Mn acceptor state may be expected. In figure
\ref{Measurements}(a) we note an increase of TMR from 30$\%$ before
annealing to 120$\%$ after annealing on a 128 $\mu$m$^2$ junction
(along [100] direction) while R.A product decreases from 0.047 to
0.003 $\Omega$.cm$^2$. Lowering R.A. product should be related to a
change in the Fermi energy which involve a reduction of the barrier
height or the barrier width and then must be associated to an
increase of the hole concentration. In addition, as already observed
on magnetic properties, the coercive field of the top magnetic layer
changes after annealing; the difference of those values between
magnetic and transport measurements is related to size effects. The
same behaviour has been observed on all 4 measured junctions [Fig.
\ref{Measurements}(b)]: Whereas TMR values lay between 30$\%$ and
90$\%$ before annealing, an homogenization of the values occurs after
annealing where TMR ranges between 110$\%$ and 130$\%$. No assymetry
of TMR between positive and
negative applied bias has been measured after annealing.\\

However, we must emphasize that magnetic properties derive from the
whole magnetic layers (volume effect), whereas electric properties
should mainly depend on the interfaces between the tunnel barrier and
the electrodes. It results that evaluating the change of the
electronic properties for each magnetic layers from transport
measurements appears more complex than in the case of magnetic
experiments. Nevertheless, some conclusions can be drawn from TMR
measurements \textit{vs.} temperature, taking into account that the
elementary process is a spin-conservative direct tunneling, i.e. the
evolution of TMR with temperature is directly linked to the effective
carrier spin polarization of the ferromagnetic
layer.\cite{MattanaPRB} We note that the effective temperature at
which TMR cancels remains unchanged after annealing
[Fig.\ref{Measurements}(c)]. This feature comes from the magnetic
properties of the bottom electrode which are not modified after
annealing (T$_C$$\sim$ 55 K). The drop of TMR around 15 K before and
after annealing is related to the quick variation of the coercive
field of the thin magnetic layer as a function of the
temperature.\cite{ChibaPhysicaE}
\newline

On the other hand, how behaves the tunnel anisotropic
magnetoresistance (TAMR)? TAMR generally traduces a variation of
resistance \textit{vs.} the crystalline orientation of the electrode
magnetization. In this case, this originates from the anisotropy of
the valence band of (Ga,Mn)As. Careful attention was paid on the
resistance difference when the magnetization is aligned along [100]
(in plane magnetization) or [001] (out of plane magnetization) which
leads to maximum TAMR effect in our samples. In a saturating field of
6 kOe variations are almost equal to 10-15$\%$ before and after
annealing [Fig. \ref{Measurements}(d)], in good agreement with
experiments obtained on a ZnSe barrier.\cite{SaitoPRL} When driving
experiments in the plane of the layer resistance variations as small
as 4$\%$ were recorded.

Combination of magnetization and transport measurements let us
therefore presume that the spin splitting and the Fermi energy play
an important role in tunneling transport. The influence of those
parameters will be discussed now through a 6x6 band k.p modelisation
of spin-orbit coupled states tunneling transport.

\section{Theoretical model}

Our calculations of the transmission coefficient are based on the
multiband transfer matrix technique developed in details by Pethukov
\textit{et al.} \cite{PethukovPRL}, Brey \textit{et
al.}\cite{BreyAPL} and Krstajic \textit{et al.}\cite{PeetersPRB} and
applied to the hole 6$\times$6 valence band \textit{k.p} Hamiltonian
$H_{h}$. Added to the Kohn-Luttinger kinetic Hamiltonian, this
includes a \textit{p-d} exchange term introduced by the interaction
between the localized Mn magnetization and the holes derived in the
mean-field approximation thus giving:

\begin{eqnarray}
H_{h}=-(\gamma_{1}+4\gamma_{2})k^{2}+6\gamma_{2}\sum_{\alpha}
L_{\alpha}^{2}k_{\alpha}^{2}+\nonumber\\
+6\gamma_{3}\sum_{\alpha\neq \beta}
(L_{\alpha}L_{\beta}+L_{\beta}L_{\alpha})k_{\alpha}k_{\beta}+\lambda_{so}\overrightarrow{L}\overrightarrow{S}+6B_G\widehat{m}\overrightarrow{S}
\label{eq1}
\end{eqnarray}
equivalent to the one proposed by Dietl \textit{et al.
}\cite{DietlPRB} and Abolfarth \textit{et al}\cite{AbolfathPRB}.
Here, $\alpha=\{x,y,z\}$, $L_{\alpha}$ are $l=1$ angular momentum
operators, $\overrightarrow{S}$ is the vectorial spin operator,
$\widehat{m}$ the unit magnetization vector and $\gamma_{i}$ are
Luttinger parameters of the host semiconductor GaAs. 6B$_G$
represents the spin-splitting between the heavy holes at the
$\Gamma_{8}$ point like originally introduced by Dietl \textit{et
al.}\cite{DietlPRB} We do not take explicitly into account the stress
hamiltonian which is shown to give the same qualitive conclusions.

To derive the transmission coefficient, the boundary conditions to
match at each interface are\cite{PethukovPRL}:

\textit{i}) the continuity of the 6 components of the envelope
function according to $\psi^{+}_{n}$+$\sum_{\overline{n}}
r_{n,\overline{n}}\psi^{-}_{\overline{n}}$=
$\sum_{n^{'}}t_{n,n^{'}}\psi^{+}_{n^{'}}$ where the subscript
$t_{n,n^{'}}$ ($r_{n,\overline{n}}$) refer to the respective
transmission (reflection) amplitude from \textit{incident} ($n$),
\textit{reflected} ($\overline{n}$) and \textit{transmitted}
($n^{'}$) waves together with

\textit{ii}) the continuity of the 6 components of the current
wavevector according to
$\widehat{J}\psi^{+}_{n}$+$\sum_{\overline{n}}r_{n,\overline{n}}\widehat{J}\psi^{-}_{\overline{n}}$=
$\sum_{n^{'}}t_{n,n^{'}}\widehat{J}\psi^{+}_{n^{'}}$ where, in the
\textit{k.p} approach, the current operator in the $z$ direction
writes $\widehat{J}=\frac{1}{\hbar}\frac{\partial H_{h}}{\partial
k_{z}}$.\newline

Concerning the heterostructure itself, the valence band offset (VBO),
d$_B$, between (Ga,Mn)As and (In,Ga)As fixes the effective barrier
height $\phi$ according to d$_B$=-E$_F$+$\phi$ where E$_F$$\sim$-0,18
eV is the Fermi level within (Ga,Mn)As calculated from the top of the
(Ga,Mn)As valence band [\textit{Inset} Fig.\ref{RA}]. On figure
\ref{RA}, we present the calculated R.A product \textit{vs.} the
respective valence band offset using standard Landauer formula of
conductance for 6 nm GaAs and In$_{0.25}$Ga$_{0.75}$As barriers.
Although the VBO between Ga$_{0.926}$Mn$_{0.074}$As and
In$_{0.25}$Ga$_{0.75}$As is still unknown, recent photoemission
spectra determined the barrier height $\phi$ between (Ga,Mn)As and
GaAs to 450 meV,\cite{AdellAPL06} in agreement with our \textit{k.p}
model considering a R.A product approaching $\sim$ 10$^{-3}$
$\Omega$.cm$^2$ [Fig. \ref{RA}] and like obtained experimentally by
Chiba \textit{et al.}.\cite{ChibaPhysicaE} The relative small band
offset between valence band of GaAs and (In,Ga)As inferior to 50 meV,
\cite{Tiwari} makes then such value of $\phi$$\simeq$ 450 meV a
plausible order of magnitude for the effective barrier height for
In$_{0.25}$Ga$_{0.75}$As matching with the R.A product after
annealing. However, in the present case, the real value of $\phi$ may
vary depending on:

\begin{figure}
\includegraphics[width=6cm]{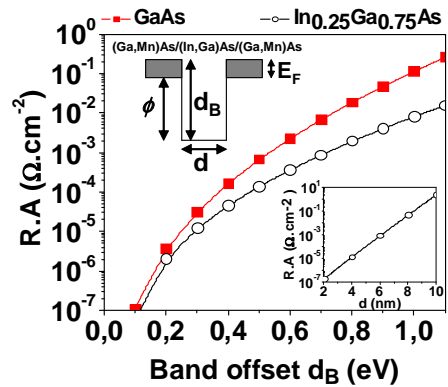}
\caption{ (Color online) Calculated Resistance.Area (R.A) product of
the trilayer structure as a function of the band offset d$_B$ between
the ferromagnetic semiconductor and a 6 nm barrier of GaAs or
(In,Ga)As. Insets : (Bottom) R.A. product as a function of the
(In,Ga)As barrier width d. (Top) Valence band profile of the
considered heterostructure. } \label{RA}
\end{figure}

\textit{i)} the nature and density of the dangling bonds at the
interfaces promoted by the low-temperature growth procedure.
\cite{LodhaaJAP}

\textit{ii)} the local density and position in energy of ionized
defects such as As antisites in the barrier which strongly influences
the valence band bending of the whole heterostructure.\newline

However, surprisingly, we have noticed that TMR remains quasi
insensitive to the barrier height $\phi$ (not shown). Let us then
focus on the phase diagrams TMR (E$_F$, B$_G$) and TAMR (E$_F$,
B$_G$) established from the preceeding model of \textit{k.p} tunnel
conduction. Figure 4 displays both TMR and TAMR \textit{vs.} the spin
splitting parameter B$_G$ and the Fermi energy E$_F$ of the
ferromagnetic semiconductor whose properties are assumed to be
identical at both sides of the (In,Ga)As barrier. The zero energy for
E$_F$ corresponds here to the top of the valence band for (Ga,Mn)As
in its paramagnetic phase (B$_G$=0). Also are plotted on phase
diagrams 3 different lines corresponding to constant carrier
concentrations of 1$\cdot$10$^{20}$ cm$^{-3}$, 3.5$\cdot$10$^{20}$
cm$^{-3}$ and 5$\cdot$10$^{20}$ cm$^{-3}$, as well as the energy of
the four first bands at the center of the Brillouin zone.\newline Let
us first emphasize on the general trends for TMR and TAMR from such
diagrams.

\section{Discussion}
High TMR values, up to several hundred percents, can be expected
either for spin splitting values larger than several tens of meV or
for low carrier concentration, that is when only the first subband is
involved in the tunnelling transport. This corresponds to a quasi
half metallic character for (Ga,Mn)As. Starting from the first
subband and increasing the carrier concentration to fill the
consecutive lower subbands (n=2,3,4), up and down spin populations
start to mix up, leading to a decrease of TMR. For high carrier
concentration (n=4), small TMR is expected which may anticipate
difficulties to conciliate high Curie temperature and large TMR
effects. We specify that for low values of spin splitting and Fermi
energy, ferromagnetic phase induced by carrier delocalization may not
exist (top right corner of the diagram) which is not taken into
account in our k.p modelisation (propagative envelope wave function).
In the same manner, we cannot reproduce metal-insulator transition in
the tunneling transport, responsible for the large TAMR obtained in
in-plane geometry.\cite{RusterPRL, PappertPRL}

\begin{figure}
\includegraphics[width=7cm]{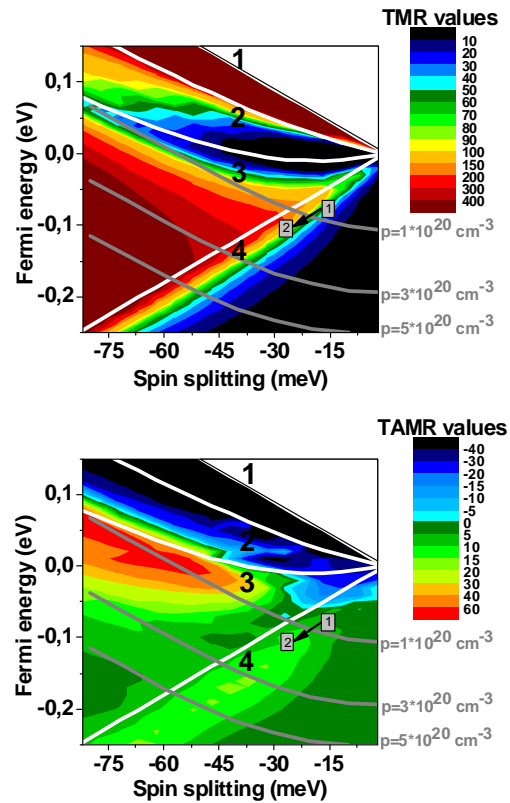}
\caption{ (Color online) Tunnel magnetoresistance values (a) and
tunnel anisotropic magnetoresistance values (b) represented as a
function of the Fermi and spin splitting energy for a 6 nm (In,Ga)As
barrier with a band offset of 450 meV. White lines represent the 4
bands at the center of the Brillouin zone. Gray lines indicate the
Fermi energy for different hole concentrations.} \label{TMRTAMR}
\end{figure}

What about TAMR signal ? We can firstly note a possible change of
sign for TAMR on crossing the third subband. The first subband
clearly gives a negative contribution to TAMR. This originates from
the predominant heavy hole character of such band, an in-plane
magnetization allowing, through off diagonal components, a possible
heavy to light hole conversion, and then a larger transmission
through the barrier.\cite{ElsenPRL} This argument is reversed for the
second and third subbands with the results that TAMR becomes positive
when n=2 and n=3 subbands are dominant in the tunneling transport. We
can point out that a change of TAMR sign was already observed on a
Zener-Esaki diode\cite{GiraudAPL} as well as theoretically
established through tight-binding treatment.\cite{SankowskiPRB}
Reducing the hole concentration through hydrogenation technique
should give the possibility to probe this possible crossover from
positive to negative TAMR.\cite{ThevenardAPL}\newline

Concerning our own experiments, taking into account conjugate TMR and
TAMR values obtained before and after annealing, one can roughly
evaluate the projection of the corresponding signals trajectories in
the [E$_F$, B$_G$] plane followed during annealing
[Fig.\ref{TMRTAMR}]. A good qualitative agreement can be found even
though symmetrical junctions were simulated in order to restrict the
number of parameters.\\

Evaluating directly the interfacial spin splitting from the mean
field theory appears difficult since the
interfacial magnetic properties are hardly accessible. However, when using the estimated
B$_G$ value of the top magnetic electrode (before and after
annealing) a good qualitative agreement can be found for TMR and TAMR, as illustrated by the trajectory in figure \ref{TMRTAMR}
between point 1 (before annealing) and point 2 (after annealing). A
more refine calculation including two different B$_G$ after annealing
should be required to draw definite quantitative conclusion.\\

We are now going to discuss the hole concentration derived from these
diagrams. TMR and TAMR values obtained before annealing are well
reproduced for a hole concentration approaching 10$^{20}$cm$^{-3}$,
in good agreement with the one measured for single (Ga,Mn)As layer
and already reported.\cite{MalfaitJMMM} The annealing procedure has
for effect to \textit{i)} remove Mn interstitial atoms, \textit{ii)}
increasing carrier concentration and  \textit{iii)} reduce the
effective barrier height even if the valence band position is
expected to rise due to an increase of the average exchange energy
(B$_G$). The large reduction of the R.A product together with the
increase of TMR are consistent with such assumption. Nevertheless,
the hole concentration extracted after annealing from the phase
diagram $\sim$1,7.10$^{20}$cm$^{-3}$ appear to be weak compared to
the one reported in the literature and derived from Hall effect
measurements. The existence of a possible concentration gradient can
be at the origin of such discrepency. Also can be invoked, a
reduction of the hole concentration at the interfaces with the
barrier due to a significant charge transfer between \textit{p}-type
(Ga,Mn)As and \textit{n}-type
(In,Ga)As (excess of As antisites).\cite{KoederAPL,PappertPRL}\\

\section{Conclusion}
In summary we have shown that annealing a (Ga,Mn)As-based tunnel
junction mainly affects the properties of the top magnetic layer,
ensuring an increase of the effective magnetization and a significant
enhancement of the tunnel magnetoresistance. The confrontation
between experiments and modelisation performed within a 6x6 band k.p
treatment \textit{vs.} intrinsic (Ga,Mn)As parameters (hole filling,
exchange energy) allowed a rough estimation of the average exchange
interactions and carrier concentration in (Ga,Mn)As at the interface
with the barrier. We point out that while the magnitude of TMR
appears very sensitive to both parameters (B$_G$ and E$_F$), the TAMR
variation is limited to several tens of percent but may change sign
crossing from upper to lower (Ga,Mn)As subbands. As a final
conclusion, we think that this reduced parameter model gives a good
qualitative agreement of the tunneling transport and enables to
extract the fundamentals of TMR and TAMR processes involving tunnel
transport of spin-orbit couple state. In order to go further and draw
more quantitative information, a perfect control
and knowledge of the carrier density seems to be necessary.\\

\begin{acknowledgements} We gratefully acknowledge H.-J. Drouhin, A.
Fert, G. Fishman and B. Vinter for fruitful discussions.

This work was supported by the EU Project NANOSPIN
FP6-2002-IST-015728 and by the french ANR Program of Nanosciences and
Nanotechnology (PNANO) project MOMES.

\end{acknowledgements}

\end {document}